\documentclass[runningheads]{llncs}
\usepackage{lmodern}
\usepackage{graphicx}
\usepackage[T1]{fontenc}
\usepackage[utf8]{inputenc}
\usepackage{amsmath}
\usepackage{booktabs}
\usepackage{array}
\usepackage{xcolor}
\usepackage{microtype}
\usepackage{hyperref}

\begin{document}

\title{Democratic Accountability for Autonomous Systems: A Legislative Notification Framework and Comparative Institutional Design}
\titlerunning{Democratic Accountability for Autonomous Systems}

\author{Samarjeet Malik\inst{1,2*} \and
Mansi Ganga Nayak\inst{3\dagger} \and
Mugdha Raje Ashiya\inst{2\dagger} \and
Hritika Adhikary\inst{3\dagger} \and
Vaibhav Sharma\inst{4}}
\authorrunning{S. Malik et al.}

\institute{Ashoka University, Sonipat, Haryana, India \and
Indian Institute of Technology Jodhpur, Rajasthan, India (Predoctoral Fellows) \and
Arizona State University, Tempe, AZ, USA \and
Cruxway, New York, NY, USA}
\maketitle
\begin{flushleft}\small
$^{*}$\,Corresponding author: Samarjeet Malik (\email{samarjeetmalik.gs5@gmail.com}).\\
$^{\dagger}$\,Mansi Ganga Nayak, Mugdha Raje Ashiya, and Hritika Adhikary contributed equally.
\end{flushleft}
\setcounter{footnote}{0}

\begin{abstract}
No domestic legislature has ever reviewed a state's decision to authorise an autonomous cyber weapon before it acted. Closing the international attribution gap is necessary but not sufficient for democratic accountability. We formalise the Tripartite Attribution Framework's three attribution levels and add a fourth, non-legal Democratic Accountability level ($A_{dem}$): a 72-hour legislative notification requirement. Scoring six democratic oversight bodies against five adoptability dimensions yields a counter-intuitive result: the statutes granting these bodies privileged access to classified material are often the same statutes withholding particular ongoing operations from them, so at most one could lawfully receive such a notification today, and the carve-out Level~4 needs is required in nearly every jurisdiction rather than only the least-equipped. Retrospective analysis of Stuxnet, NotPetya, and Volt Typhoon shows one pattern: legal attribution resolvable, domestic democratic review absent.

\keywords{democratic accountability \and AI governance \and legislative oversight \and classified-oversight institutions \and autonomous cyber weapons \and command responsibility \and parliamentary notification}
\end{abstract}

\section{Introduction}

When algorithmic systems make engagement-relevant decisions, oversight typically terminates inside classified command structures, never reaching the domestically accountable institutions (legislatures, oversight bodies, the public) that are supposed to constrain a state's use of force on its own citizens' behalf.

This is structurally the same governance problem the EU AI Act \cite{aiact2024}, the Digital Services Act \cite{dsa2022}, and the UK Online Safety Act \cite{uk_osa2023} solve for commercial platforms (assign responsibility, require oversight, preserve a public record), except that each instrument excludes state actors (Art.~2(2) DSA; AI Act Recital~27). This paper asks what that answer looks like when the deployer is a state, not a company.

\noindent\textbf{Where this sits in the conference's questions.} The 2026 call asks how AI systems and platforms are governed today, who is responsible for mitigating digital harms, and what policy frameworks can balance innovation, accountability and public trust. This paper answers the second and third. Social cybersecurity treats cyber operations as acting on societies through the institutions that mediate them \cite{carley2020}; the institutions examined here are those a democracy uses to hold its own state to account for such operations, and our finding is that they are at present statutorily unable to perform that function for autonomous systems. Identifying that gap is the task the call names first.

The pattern recurs across three attributed cases, developed formally in Section~4: no congressional committee was briefed on Stuxnet's authorisation before deployment \cite{sanger2012}; no Russian legislative body reviewed NotPetya's self-propagating capability before its estimated \$10 billion in damage across 65 countries \cite{greenberg2019}; and no oversight committee reviewed China's authorisation of Volt Typhoon's AI-driven lateral movement inside U.S.\ infrastructure \cite{cisa2024}, nor, as Section~6 argues, could one plausibly have done so, since no Chinese institution holds a comparable mandate.

This paper makes four contributions: (1)~a formal restatement of TAF's three attribution levels, each anchored to a doctrinal test (Tallinn Manual Rule~15, ARSIWA Art.~4, Rome Statute Art.~28) with worked examples from the three case studies; (2)~a 72-hour legislative notification requirement for deployments that have cleared Level~3 (command) attribution, enforced through a compliance-cost asymmetry between substantive and performative review; (3)~a comparative institutional analysis of six democratic oversight bodies (the UK ISC, US congressional intelligence committees, India's SCOD, Germany's PKGr, France's DPR, and Australia's PJCIS) scored against five adoptability dimensions, yielding the counter-intuitive finding that statutory access and operational review are traded off against one another, so the best-equipped bodies are the ones most clearly barred; and (4)~a retrospective case-study analysis showing the framework would have created a domestic accountability opportunity in three major cyber operations where none existed.

\section{Background: The Democratic Oversight Gap}

\noindent\textbf{The IHL Context.} Existing IHL assumes a human decision-maker at the point of attack: the Tallinn Manual 2.0 \cite{schmitt2017} addresses attribution (Rules~14--17) and defines a cyber attack by its effects (Rule~92), both on the premise that a human pulled the trigger, and Additional Protocol I's Article~36 obliges states to review new weapons for IHL compliance, though no state has publicly disclosed an Article~36 review of an AI-driven cyber tool. An autonomous proximate actor opens two doctrinal gaps: a \emph{mens rea} gap (the system executes the attack but cannot form criminal intent, the accountability gap Crootof's ``war torts'' proposal identifies for autonomous weapons and answers with state, rather than individual, responsibility \cite{crootof2016wartorts}) and a \emph{temporal displacement} gap (the authorisation decision and the attack's execution are separated in time, letting a commander disclaim knowledge of the system's specific actions).

Existing doctrine closes the first gap and, importantly, does \emph{not} close the second. On attribution, the right anchor is ARSIWA Article~4 rather than Article~8: the conduct of a State organ is attributable to the State as such, so where an autonomous system is fielded by a military cyber command the attribution question does not turn on who or what executed the action. Article~8, which the literature more often reaches for, governs \emph{non-State} actors and imports the ICJ's \emph{Nicaragua} effective-control test, under which even financing, organising, equipping, selecting targets and planning an operation as a whole was held insufficient \cite{arsiwa2001,icj1986}. That test is exacting and operation-specific; invoking it for a State's own organ would make the argument harder, not easier. Relocating responsibility to the authorisation decision therefore works, but as an Article~4 point about organs, not an Article~8 point about control.

The \emph{mens rea} gap is more stubborn, and the leading authority runs against the intuitive fix. It is tempting to argue that a commander who knows an autonomous system's documented capabilities thereby ``should have known'' its output, satisfying Rome Statute Article~28. The ICTY Appeals Chamber in \emph{Bla\v{s}ki\'{c}} foreclosed the equivalent move under Article~7(3): it reversed the Trial Chamber for reading ``had reason to know'' as ``should have known'', rejected a negligence standard outright, and held that liability requires information actually available to the superior putting him on notice of offences, since ``the knowledge of any kind of risk, however low, does not suffice'' \cite{blaskic2004}. Knowing what a system \emph{could} do is precisely the showing that was held inadequate. Rome Statute Article~28(a)(i) does set a deliberately broader standard for military commanders (``knew or, owing to the circumstances at the time, should have known''), so the argument is not foreclosed in the ICC's own terms \cite{rome1998}; but it cannot borrow \emph{Bla\v{s}ki\'{c}}'s authority, and the gap it must bridge is wider than the analogy suggests. An autonomous system acting at machine speed may generate no information putting a commander on notice before the harm occurs. That is the accountability deficit this paper addresses, and command-responsibility doctrine deepens rather than resolves it.

This relocation resolves only inter-state legal responsibility; it does not require the authorisation record to ever reach a domestic legislature, an independent oversight body, or the public. A state can in principle satisfy every evidentiary requirement of international legal attribution internally while disclosing none of it domestically, the gap this paper addresses.

\noindent\textbf{Existing Oversight Institutions.} Democratic states already maintain classified-oversight bodies that receive sensitive operational material unsuited to the public record; the comparative legal standards for such bodies are well developed in the intelligence-accountability literature \cite{bornleigh2005}. The question is whether their statutory mandate extends to AI cyber deployment-authorisation records and, where it does not, what would be required to extend it.

Section~6 scores six such bodies (Table~\ref{tab:institutions}), and the result is counter-intuitive: statutory access to classified material does not imply the power to receive a report on a \emph{particular} operation, and in several jurisdictions the same instrument confers the first and withholds the second. Section~5 specifies a mechanism that assumes, rather than invents, the oversight architecture it needs.

\section{The Tripartite Attribution Framework and Democratic Accountability}

The Tripartite Attribution Framework (TAF)\footnote{This paper's contribution is institutional design (who receives Level~4 evidence, under what statutory authority, and on what clock) rather than attribution methodology. TAF's three legal levels are restated here only to the depth the Level~4 argument requires; the paper does not claim to advance the evidentiary standards themselves.} closes an inter-state legal gap, giving international tribunals, treaty bodies, and other states an evidentiary basis for assigning responsibility for an AI-initiated cyber operation at three independent levels: which state's system acted (technical), who deployed it (operational), and who commanded the deployment (command); a finding at one level does not entail a finding at another. Formally, $A_{total} = (A_{tech}, A_{op}, A_{cmd})$. But, as Section~2 argues, international legal accountability is not the only accountability deficit autonomous weapons raise.

\noindent\textbf{Formal Definitions and Worked Examples.} We formalise each TAF level as a doctrinal test, then anchor it to a worked example drawn from one of the three case studies in Section~4.

\noindent\textbf{Definition 1 (Technical Attribution, $A_{tech}$).} A state $S$ satisfies $A_{tech}$ for an operation $\mathcal{O}$ when forensic evidence (code and payload fingerprints, infrastructure provenance, timing and key-rotation signatures) establishes that a system under $S$'s effective control executed $\mathcal{O}$. The doctrinal anchor is Tallinn Manual 2.0 Rule~15, which attributes to a State the cyber operations of its organs, together with its commentary on evidentiary sufficiency: the Manual is explicit that an operation merely originating from, or reporting back to, governmental cyber infrastructure is ``usually insufficient evidence for attributing the operation to that State'' \cite{schmitt2017}. We therefore read the threshold as a \emph{preponderance of independent forensic markers converging on} $S$, no one of which would suffice alone, which is precisely the structure of the NotPetya evidence below.

\noindent\textit{Worked example (NotPetya).} $A_{tech}$ is satisfied for Russia by the forensic markers in the October 2020 U.S. DOJ indictment of six GRU officers: reuse of the Mimikatz credential-theft routine, KillDisk wiper logic reused from prior Sandworm operations, and infrastructure overlap with the 2016 WADA intrusion (paras.~24--31, 38--42) \cite{doj2020}.

\noindent\textbf{Definition 2 (Operational Attribution, $A_{op}$).} $A_{op}$ is satisfied when a documented deployment-authorisation record identifies the unit or commander who configured and activated the system within defined parameters, treating the act of deployment itself as a triggering act of state responsibility under ARSIWA Article~4, the conduct of a State organ being attributable to the State as such \cite{arsiwa2001}.

\noindent\textit{Worked example (Stuxnet).} $A_{op}$ remains unsatisfied: no public deployment\-/\-authorisation record has surfaced; the only documentary trace is the operation's reconstruction by Sanger \cite{sanger2012} and Zetter \cite{zetter2014}, neither disclosing an authorising unit or commander.

\noindent\textbf{Definition 3 (Command Attribution, $A_{cmd}$).} $A_{cmd}$ is satisfied when the authorising commander knew or should have known the category of acts the system would carry out and failed to take reasonable measures to prevent their unlawful use, satisfying the Rome Statute Article~28 standard \cite{rome1998,blaskic2004}.

\noindent\textit{Worked example (Volt Typhoon).} $A_{cmd}$ is unresolvable against China: the MSS--PLA command chain for offensive cyber operations is undocumented in any public record, and the CISA advisory \cite{cisa2024} identifies the actor but not an authorising commander, leaving the Article~28 test unsatisfiable on available evidence.

\begin{figure}[h]
\centering
\includegraphics[width=0.9\textwidth]{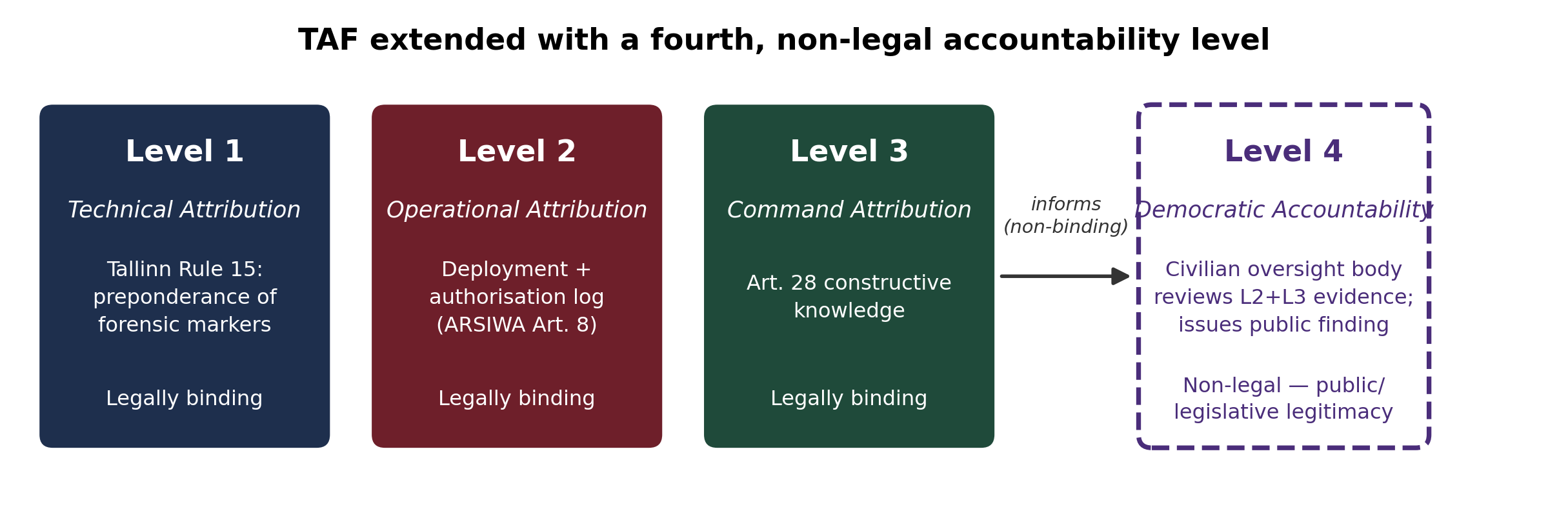}
\caption{TAF extended with Level 4 (Democratic Accountability). Levels 1--3 remain the legally binding core; Level 4 is a parallel, non-binding layer that consumes Level 2/3 evidence to produce a public accountability finding.}
\label{fig:taf-l4}
\end{figure}

\noindent\textbf{Level 4, Democratic Accountability.} Levels~1--3 establish legal responsibility: technical attribution sufficient for international fact-finding, operational attribution sufficient to identify the deploying unit, and command attribution sufficient for individual criminal liability. None requires disclosure to any domestic body, which is the gap identified in Section~2.

We therefore extend TAF with a fourth, non-legal level (Fig.~\ref{fig:taf-l4}). Democratic Accountability ($A_{dem}$) is established when a statutorily empowered civilian oversight body reviews the Level~2 and Level~3 evidence and issues a public accountability finding, redacted where classification requires. It is deliberately not a precondition for criminal liability under Levels~1--3; its function is legitimacy rather than punishment, converting TAF's evidence chain into a form a domestic public can evaluate. Section~5 operationalises it as a 72-hour notification requirement, and Section~6 asks which institutions could receive one.

\section{Case Studies: Democratic Failures in Retrospect}

Applying TAF retrospectively to three attributed operations shows the same pattern whatever the strength of the underlying legal case.

\noindent\textbf{Stuxnet (2010).} Technical attribution is strong (exploit signatures, zero-days, certificate chains), with US--Israeli origin confirmed via the 2013 Snowden disclosures \cite{zetter2014}; operational and command attribution remain unsatisfied, no public authorisation record existing. The democratic failure is independent of both: no congressional committee was briefed on Operation Olympic Games before deployment, and disclosure came years later through journalism \cite{sanger2012}.

\noindent\textbf{NotPetya (2017).} Level~1 is resolved by the 2020 DOJ indictment of Sandworm \cite{doj2020}, but the chain authorising autonomous propagation beyond Ukraine remains undocumented at Levels~2 and 3. No Russian legislative body reviewed a capability that caused an estimated \$10 billion in damage across 65 countries \cite{greenberg2019}; sanctions substitute for international consequence, not for domestic review in the authorising state.

\noindent\textbf{Volt Typhoon (2024).} The most AI-intensive case has the weakest Level~1 attribution, ``living off the land'' techniques deliberately avoiding TTP fingerprints, though CISA attributed the campaign to China \cite{cisa2024}. Here the failure is structural: no Chinese institution holds a mandate comparable to the ISC or HPSCI/SSCI, so Level~4 has nothing to attach to.

\begin{figure}[t]
\centering
\includegraphics[width=\textwidth]{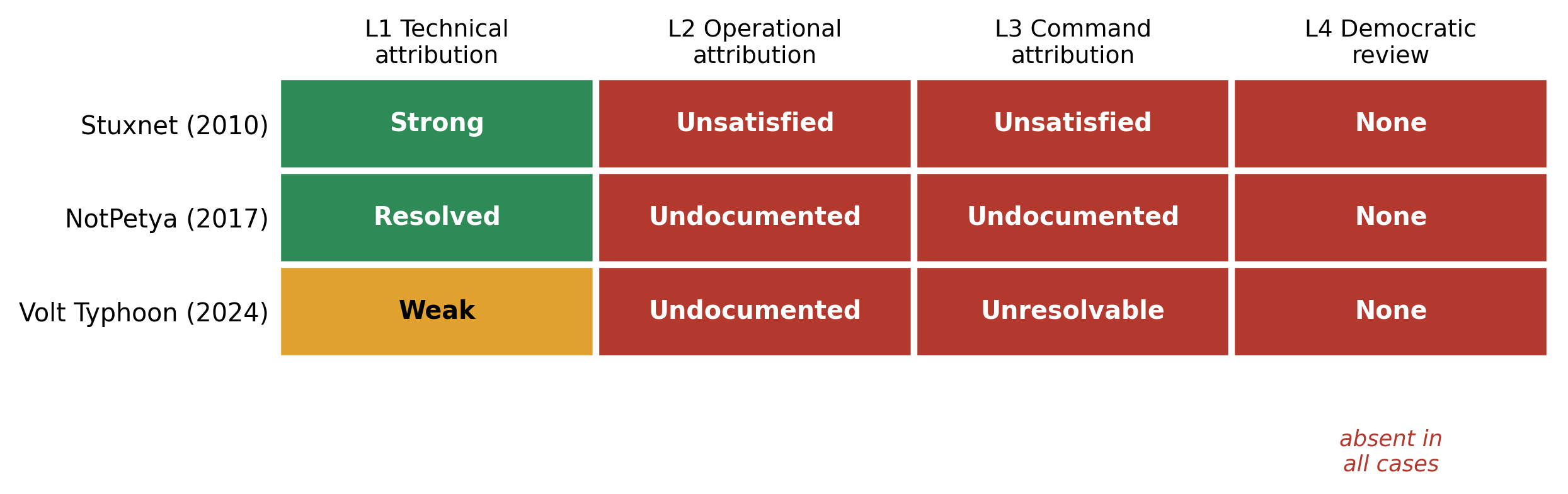}
\caption{The three case studies against TAF's four levels. Technical attribution varies from strong to weak and the legal levels are largely unsatisfied on the public record, but Level~4 review is absent in every case: the invariant this paper targets.}
\label{fig:case-pattern}
\end{figure}

One pattern holds across all three (Fig.~\ref{fig:case-pattern}): technical attribution ranges from strong to weak, legal attribution is eventually resolvable in two of three, and domestic democratic accountability is zero in all three. The variable that does not vary is Level~4.

\section{The Legislative Notification Framework}

\subsection{Trigger Conditions and Notification Requirements}

The notification requirement applies to deployment of any autonomous system that has cleared Level~3 attribution (a system capable of initiating cyber operations without real-time human approval of the individual engagement) for offensive operations with potential civilian-infrastructure effects, explicitly excluding purely defensive AI tools (addressing the chilling-effect objection in Section~7). Two senses of ``cleared'' must be kept apart: Section~4 assesses what \emph{external} observers could establish after the fact, whereas the trigger here is the deploying state's own \emph{internal} determination at authorisation, which is always available to it. That asymmetry is why domestic notification can operate where international attribution stalls.

The recipient is a designated legislative oversight body with a statutory classified-access mandate (e.g., the ISC or HPSCI/SSCI), within 72 hours of deployment authorisation. Required content: (i)~confirmation that TAF Levels~2 and 3 have been internally established; (ii)~confirmation that a proportionality assessment was conducted; and (iii)~identification of the target scope and foreseeable collateral effects, not the underlying classified evidence itself. The output is a redacted summary, publishable under standard classification review and modelled on existing ISC practice.

\noindent\textbf{Legal Anchor and Feasibility.} The requirement rests on domestic legislative-oversight statute rather than international law: it needs no treaty and no other state's cooperation, and is adoptable unilaterally, being closer to an existing classified-briefing obligation than to a new international commitment. Adoptability nonetheless varies sharply by jurisdiction, and Section~6 shows the variation runs the opposite way to the one the literature would predict.

\begin{figure}[h]
\centering
\includegraphics[width=\textwidth]{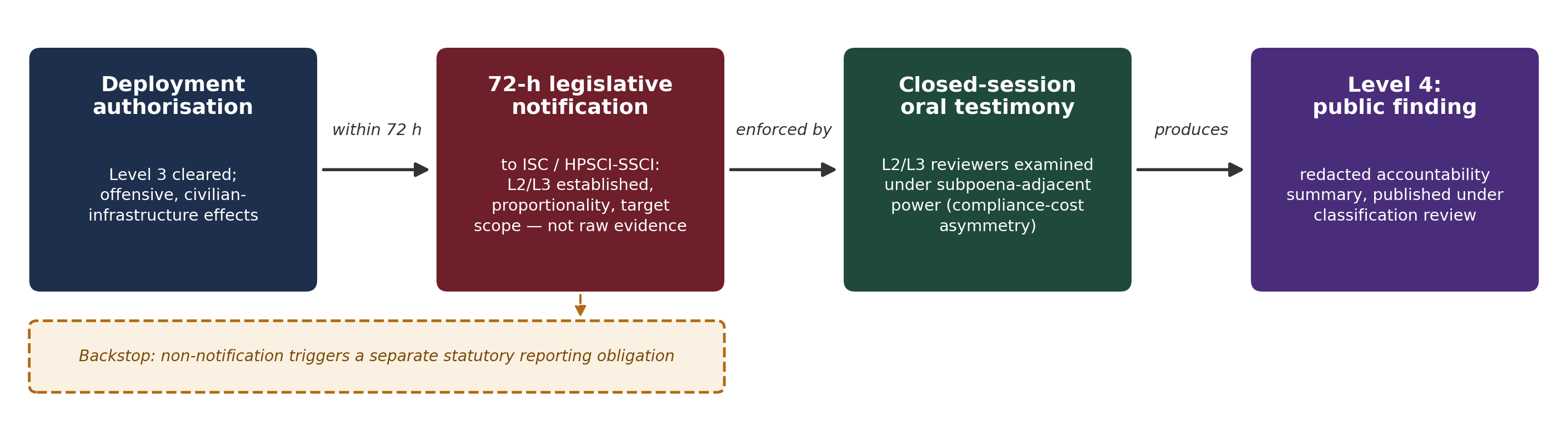}
\caption{The 72-hour legislative-notification mechanism (Section~5). A Level-3-cleared deployment triggers a procedural notification within 72 hours; closed-session oral testimony of the reviewers enforces it through the compliance-cost asymmetry; the output is a redacted Level-4 public accountability finding. Non-notification triggers a separate statutory reporting obligation as a backstop.}
\label{fig:notification-flow}
\end{figure}

\noindent\textbf{Enforcement: The Compliance-Cost Asymmetry.} A notification requirement satisfiable with a written summary is vulnerable to hollow compliance (Fig.~\ref{fig:notification-flow}). The framework addresses this with a procedural design borrowed from ISC practice under the Justice and Security Act 2013: the oversight body must have standing authority to interview, in closed session, the technical personnel who conducted the Level~2/3 review \cite{isc2013}.

This produces a compliance-cost asymmetry: written notification is cheap to fabricate, but oral testimony under subpoena-adjacent power is expensive to fabricate consistently, since follow-up questioning can expose whether the named reviewer examined the authorisation record, what proportionality factors were weighed, and at what command level the review was approved. Sustaining false testimony across rounds costs more than the political price of not notifying, particularly once non-notification triggers a separate reporting obligation. The mechanism must be classified honestly, because it cuts against the oversight literature. McCubbins and Schwartz distinguish costly, centralised ``police-patrol'' oversight from cheap, decentralised ``fire-alarm'' oversight, and argue legislators rationally prefer alarms precisely because patrols are expensive \cite{mccubbins1984}. Closed-session examination of review personnel is unambiguously a patrol, and the literature predicts under-investment in exactly this. Our answer is that the 72-hour trigger changes the economics rather than the mode: an unbounded duty to go looking becomes a bounded duty to respond to a dated notification, so the notification performs the alarm's function while the examination retains the patrol's verification power. Whether that hybrid survives contact with legislative incentives we do not settle; we are aware of no direct test of the asymmetry in the classified-oversight setting, and the claim is a design inference.

\section{Comparative Institutional Design}

This comparative analysis identifies which oversight functions already exist in domestic statute and which states would need to build them from scratch.

\noindent\textbf{Statutory Architecture: UK, US, India.} The UK ISC, established by the Intelligence Services Act 1994 and strengthened by the Justice and Security Act 2013, has statutory access to highly classified material: section~2 of the 2013 Act inserts section~10A into the 1994 Act, giving the Committee power to require information from the intelligence agencies. Its precise remit (including oversight of the National Cyber Force) is set out in the Memorandum of Understanding between the Prime Minister and the ISC, revised when the NCF was established in 2020 \cite{isc2013,iscmou}: the Act establishes the ISC's \emph{power}, the MoU its \emph{remit}. Evidence flows to the ISC in closed session under the Official Secrets Act 1989, with the Committee empowered to request Level~2/3 documentation directly and publish a redacted summary subject to Prime Ministerial review.

The United States affords a close statutory analogue in the tiered congressional notification structure under 50 U.S.C. \S 3093: covert actions are ordinarily briefed to the full HPSCI and SSCI, with an ``extraordinary circumstances'' exception restricting initial notification to the Gang of Eight (the four congressional leaders plus the HPSCI/SSCI chairs and ranking members), provided the full committees are ``fully informed'' within a bounded period \cite{belfer2009,usc3093}, the model for Section~5.1's 72-hour, redacted-summary structure. That this structure exists does not make it effective: the scholarship on congressional intelligence oversight documents persistent structural weaknesses in how these committees actually exercise it \cite{zegart2011}, which is precisely why our design leans on the oral-testimony enforcement mechanism rather than notification alone.

India's Parliamentary Standing Committee on Defence (SCOD), a 31-member committee overseeing the Ministry of Defence \cite{scod2026}, illustrates the opposite case: it has no statutory right to classified operational briefings comparable to the ISC's or HPSCI/SSCI's, a scope condition under which domestic adoption would first require establishing that legal authority.

\subsection{A Complication: Title 10 vs.\ Title 50}

The U.S.\ architecture described above presumes AI-driven offensive cyber operations would be briefed under \S 3093's covert-action framework, but the U.S.\ operates offensive cyber under two parallel authorities: Title~50 (intelligence activities, including covert action under \S 3093) and Title~10 (military activities conducted by U.S. Cyber Command), with a contested boundary between covert action and ``traditional military activity'' that has generated a sustained national-security-law debate \cite{chesney2012,dodd3020}.

An AI-driven operation conducted under Title~10 is not automatically captured by \S 3093's Gang-of-Eight regime, since the ``traditional military activity'' exemption in \S 3093(e)(2) excludes operations under military command in anticipated hostilities \cite{usc3093}. Title~10 supplies its own, quite different reporting regime: 10 U.S.C. \S 394 authorises military cyber operations ``including clandestine military activities or operations in cyberspace'' and routes their reporting to \S 484, which requires only that senior officials ``provide to the congressional defense committees \emph{quarterly} briefings on all offensive and significant defensive military operations in cyberspace, including clandestine cyber activities, carried out by the Department of Defense during the immediately preceding quarter'' \cite{usc394,usc484}.

\noindent\textbf{An asymmetry at the heart of U.S.\ practice.} Set that against 10 U.S.C. \S 130f, which obliges the Secretary of Defense to notify the congressional defense committees \emph{in writing within 48 hours} of any sensitive military operation, a category built around lethal or capture operations directed at specific individuals \cite{usc130f}. U.S.\ law therefore treats a single kinetic operation against a named person as requiring notice within two days, while an entire quarter of offensive and clandestine cyber activity is discharged by one aggregate briefing. A quarterly aggregate cannot, by construction, surface the authorisation of any particular autonomous deployment: the reporting unit is the quarter, not the operation. This is the Title~10 half of the accountability gap, and it is visible on the face of the statutes rather than inferred.

Two consequences follow. First, the framework's correct U.S.\ anchor is the combination of \S 3093 (Title~50, covert action, HPSCI/SSCI) and \S\S 394 and 484 (Title~10, military cyber, congressional \emph{defense} committees). The seam is therefore not only an authorities question but a committee-jurisdiction split: the same AI-driven operation reaches different overseers depending on which title it is run under, and neither overseer sees a per-deployment record. Second, the 72-hour requirement of Section~5.1 is a tightening of an existing duty rather than a new imposition. Congress has already legislated a hard 48-hour clock for sensitive operations under \S 130f and a standing cyber-reporting duty under \S 484; Level~4 asks that the cyber duty be discharged per deployment rather than per quarter, which is adoptable by amending \S 484 rather than by building new machinery. This is the paper's position on a contested question in the national-security-law literature \cite{chesney2012}, not a settled one.

\noindent\textbf{A unified six-body comparison.} Germany's Parlamentarisches Kontrollgremium (PKGr), France's D\'el\'egation parlementaire au renseignement (DPR) and Australia's Parliamentary Joint Committee on Intelligence and Security (PJCIS) join the UK, US and Indian bodies analysed above in Table~\ref{tab:institutions}.

\begin{table}[h]
\centering
\caption{Six democratic oversight bodies scored against the framework's five adoptability dimensions.}
\label{tab:institutions}
\footnotesize
\setlength{\tabcolsep}{3.2pt}
\newcolumntype{L}[1]{>{\raggedright\arraybackslash}p{#1}}
\begin{tabular}{L{2.0cm}L{1.85cm}L{1.45cm}L{1.25cm}L{1.8cm}L{2.45cm}}
\toprule
\textbf{Institution} & \textbf{Stat. access} & \textbf{Compell.} & \textbf{Closed session} & \textbf{Public finding} & \textbf{AI/cyber mandate} \\
\midrule
UK: ISC \cite{isc2013,iscmou} & Yes (JSA 2013 s.~2) & Yes & Yes & Yes (redacted, PM review) & NCF via MoU, but s.~2(3) bars ongoing ops \\
US: HPSCI/SSCI \cite{usc3093,usc394,usc484} & Yes (50 U.S.C. \S~3093) & Yes & Yes & Limited (classified) & Partial (Title 10/50, \S~6.1) \\
India: SCOD \cite{scod2026} & No (voluntary only) & No & No & No & No \\
Germany: PKGr \cite{pkggrg2009} & Duty to inform, refusable & No & Yes (in camera) & Limited (classified only) & No: BND/BfV/MAD only, not Bundeswehr cyber \\
France: DPR \cite{dpr2007,dpr2015} & Yes (Loi\allowbreak\ 2007-1443) & Limited & Yes & Limited (annual report) & Art.~6 nonies bars ongoing ops and authorising instructions \\
Australia: PJCIS \cite{pjcis2001} & Yes (IS Act 2001 s.~29) & Yes & Yes & Yes (tabled reports) & ASD in remit, but s.~29(3) bars particular ops \\
\bottomrule
\end{tabular}
\end{table}

Five of the six meet the statutory-access and closed-session prerequisites; only India's SCOD fails them outright. Read as a readiness ranking, that is an encouraging result, and it is the reading we initially took. It is wrong, and the way it fails is the paper's central finding.

\subsection{The Prerequisite Trap}

The five dimensions of Table~\ref{tab:institutions} measure whether a body can receive classified material \emph{in general}. A Level~4 notification is not general. It is a report, within 72 hours of authorisation, identifying the target scope and foreseeable collateral effects of a deployment that is about to run or is already running. It is therefore, by construction, a \emph{particular operation, ongoing or proposed} — and that is the precise category the enabling statutes carve out.

The two bodies that score highest are the two most clearly barred. The UK ISC may consider a particular operational matter only insofar as the Committee \emph{and the Prime Minister} are satisfied that it ``is not part of any ongoing intelligence or security operation'' (JSA 2013 s.~2(3)), with any such consideration further constrained by the Prime Ministerial MoU (s.~2(4)) \cite{isc2013,iscmou}. Two independent bars apply: the notification concerns an ongoing operation by definition, and the gatekeeper is the head of the executive the notification exists to hold accountable. Australia's PJCIS is barred more flatly still: IS Act 2001 (Cth) s.~29(3) excludes from its functions the review of ``particular operations that have been, are being or are proposed to be undertaken'' by the agencies it oversees, including the Australian Signals Directorate, the very agency that would conduct an offensive cyber deployment \cite{pjcis2001}. Australia does not partially qualify; it is excluded by the same statute that grants the access.

The generalisation is uncomfortable and, we think, correct: \emph{the statutory provisions that grant these bodies privileged access to classified material are, in several jurisdictions, the same provisions that withhold particular operations from them.} Access and operational review are traded off against each other by design, because the bargain that made classified access politically acceptable was a promise not to second-guess operations in flight. Level~4 asks for exactly what that bargain excluded. On the evidence assembled here the honest count of bodies able to receive a Level~4 notification under existing statute is not five but at most one: the US congressional intelligence committees, whose \S~3093 covert-action framework is prospective by design, and even there only for operations run under Title~50 rather than Title~10 (Section~6.1).

\noindent\textbf{Five barriers, three mechanisms.} The exclusion is not uniform in form, and the differences matter for what reform would cost. Australia's is \emph{categorical}: s.~29(3) removes particular operations from the Committee's functions outright. The UK's is a \emph{dual gate}: the ISC may reach such a matter only where the Committee and the Prime Minister concur it is not part of an ongoing operation (JSA 2013 s.~2(3)), so the executive vetoes scrutiny of itself. France's is the widest: article~6 nonies withholds from the DPR not only ``les op\'erations en cours'' but ``les instructions donn\'ees par les pouvoirs publics \`a cet \'egard'' \cite{dpr2007}, which is the authorisation record itself, precisely the document a Level~4 notification transmits. Canada's is a \emph{reasoned ministerial veto}: NSICOP may review any national-security activity unless it is ongoing and the Minister determines review would be injurious, in which case reasons must be given and the bar lifted when the operation ends (s.~8(1)(b)--(3)) \cite{nsicop2017}. Germany fails earlier: the PKGr's remit runs to BND, BfV and MAD, not the Bundeswehr's cyber forces, so a military deployment falls outside its jurisdiction before any carve-out is reached, and the PKGrG structures oversight as a \emph{duty to inform} the executive may decline where the \emph{Kernbereich exekutiver Eigenverantwortung} is engaged \cite{pkggrg2009}. Three mechanisms, then: categorical exclusion, executive veto with or without reasons, and absence of jurisdiction. Germany is the partial case rather than a flat bar, its obstacle jurisdictional and discretionary rather than categorical. Canada already contains, in embryo, the machinery Level~4 needs: an exclusion that is bounded, reasoned and time-limited rather than absolute.

That gradation supplies the reform template. A workable carve-out need not abolish the ongoing-operations exclusion; it needs to make it look like Canada's rather than Australia's, and then add one element none of the three has: a duty to notify \emph{at authorisation} rather than a discretion to review \emph{afterwards}. Concretely, a designated body should receive the deployment-authorisation record for an autonomous system within 72 hours notwithstanding the general exclusion; the responsible minister may defer the Committee's substantive review while the operation runs, but must give reasons for the deferral and must lift it when the operation ends. That preserves the operational-secrecy bargain the exclusions were built to protect while ensuring the record exists, is held by someone outside the executive, and becomes reviewable on a defined trigger rather than never.

This reverses the paper's practical conclusion without weakening it. India's SCOD is not the outlier requiring reform; it is the honest baseline. Level~4 requires statutory amendment in every jurisdiction studied, but a narrow and patterned one, and the Canadian provisions show it is drafted already in all but the notification limb.

\noindent\textbf{A coding-reproducibility probe.} Every cell derives from the statutes cited \cite{isc2013,iscmou,usc3093,pjcis2001,scod2026,pkggrg2009,dpr2007}, but the coding remains ours, so we tested whether it reproduces. Three instances of a large language model, each in a separate context with no access to this paper or to one another, coded all six bodies against the five dimensions plus an explicit sixth (may the body lawfully receive material on a \emph{particular} ongoing or proposed operation?), working from the named instruments. Agreement was substantial: Fleiss $\kappa = 0.82$ across 36 cells, unanimous on 30, perfect on statutory access, closed session, public finding and the sixth dimension, and poor on compellability ($\kappa = 0.24$), which we flag as the least stable column. Against our coding they agreed exactly on 53\% of cells and within one ordinal step on 94\%, differing systematically by reading several of our ``Yes'' cells as ``Partial''; our grid is the more generous. The finding that matters reproduced unanimously: all three placed the UK, India, France and Australia at ``No'', the United States at ``Yes'', and Germany at ``Partial''. That differs from ours in two cells and reaches the same conclusion by an independent route, exactly one of six able to receive a Level~4 notification today. This was an automated probe, not a second human coder, and does not substitute for one.

All six sets of cells derive from the instruments cited rather than from secondary commentary. The German and French readings turn on provisions we have not seen applied to a military cyber deployment, since none has been attempted, and a regional specialist may characterise the PKGr's remit over the Bundeswehr differently than we do. We note also that the coding in Table~\ref{tab:institutions} was performed by the authors against the five definitions given above rather than by independent coders, and report it as a documented reading of primary sources rather than as a measured inter-coder result.

\section{Objections and Responses}

\noindent\textbf{Objection 1, Sovereignty and classification.} States will resist a new disclosure obligation as an infringement on military secrecy, and Level~2/3 evidence may rest on classified material unsuited to any public record. \emph{Response.} Notification requires only confirmation that internal review occurred and a proportionality conclusion, not the underlying evidence; existing \emph{in camera} practice (ICJ classified-evidence protocols \cite{icj1986} and ISC closed sessions) already handles material of this kind, under the same redaction model the ISC applies today.

\noindent\textbf{Objection 2, Chilling effect on defensive AI.} A new disclosure obligation may deter states from deploying AI defensively. \emph{Response.} The framework applies only to offensive operations with civilian-harm potential; defensive AI tools are explicitly excluded (Section~5.1).

\noindent\textbf{Objection 3, Withheld establishment.} Section~5's asymmetry answers a \emph{perfunctory} notification, but not a simpler evasion: since the trigger is the state's own internal determination that Level~3 is established, an executive can decline to make that determination at all and never trigger the duty. \emph{Response.} This is the mechanism's real vulnerability and it cannot be closed from inside the notification requirement. It has to be closed by drafting: the obligation must attach to the \emph{authorisation} of a qualifying deployment rather than to a subsequent internal finding, so that the record's existence is not discretionary. That is why the carve-out proposed in Section~6.2 specifies notification at authorisation, and why the Canadian model matters, since it already pairs an exclusion with a duty to give reasons for invoking it.

\section{Conclusion}

TAF's evidentiary chain is not useful only to international tribunals. Levels~2 and 3 (deployment authorisation and command authorisation) are exactly the records a domestic legislature needs to ask whether its own state's military AI use was proportionate and properly authorised. Level~4 and the Section~5 notification framework make that chain available domestically as well as internationally, a transparency architecture for legislatures, civil society, and the public conscience invoked by the Martens Clause.

\noindent Mapping onto the IDeaS 2026 call's three questions: (i)~\emph{how AI systems are governed today}, Section~6 shows that six democratic states possess oversight bodies with privileged access to classified material, yet that the statutes conferring that access generally bar them from the particular ongoing operations a Level~4 notification concerns, so the capability is narrower than it appears; (ii)~\emph{who is responsible for mitigating digital harms}, Level~4 relocates responsibility from the executing system to the authorising commander (Rome Statute Art.~28) and the notifying executive (72-hour requirement), paralleling the DSA's assign-responsibility logic; (iii)~\emph{what policy frameworks balance innovation, accountability, and public trust}, excluding defensive AI tools (Section~5.1) and remaining non-binding and parallel to criminal liability preserves innovation space while creating a public-record architecture that converts internal review into a form the public can evaluate.

The case studies in Section~4 trace rising operational autonomy: from a human-supervised deployment (Stuxnet) \cite{sanger2012} to AI-driven techniques designed to evade the fingerprints a human reviewer would need after the fact (Volt Typhoon) \cite{cisa2024}. As offensive AI automation and defensive architectures both keep advancing \cite{nayak2026futuretrends}, this oversight window will keep narrowing, and any notification framework will need periodic revisiting to track the underlying technology. The question is whether states accept accountability before the next NotPetya, or only after.

\begin{credits}
\subsubsection{\discintname}
The authors have no competing interests to declare that are relevant to the content of this article.
\end{credits}

\bibliographystyle{splncs04}
\bibliography{references}

\end{document}